\documentclass[10pt, conference, letterpaper]{IEEEtran}

\usepackage[utf8]{inputenc}
\usepackage{url}
\usepackage{amssymb}
\usepackage{amsmath}
\usepackage{amsthm}
\usepackage{amsfonts}
\usepackage{mathrsfs}
\usepackage{authblk}
\let\labelindent\relax
\usepackage[inline]{enumitem}
\usepackage{soul}
\usepackage{todonotes}
\usepackage{balance}

\usepackage{cite}
\usepackage{algorithmic}
\usepackage{graphicx}
\usepackage{textcomp}
\usepackage{xcolor}

\usepackage[utf8]{inputenc}

\theoremstyle{definition} \newtheorem{definition}{Definition}

\theoremstyle{remark} \newtheorem*{remark}{Remark}

\theoremstyle{plain} 

\theoremstyle{remark} \newtheorem*{ir}{IR}

\begin{document}

\title{\textbf{Merkle-CRDTs}\\ Merkle-DAGs meet CRDTs}

\author[1]{H\'{e}ctor Sanju\'{a}n} \author[2]{Samuli P\"{o}yht\"{a}ri} \author[1]{Pedro Teixeira} \author[1,3]{Ioannis Psaras}

\affil[1]{Protocol Labs}
\affil[2]{Haja Networks}
\affil[3]{University College London}

\maketitle

\begin{abstract}

We study Merkle-DAGs as a transport and persistence layer for Conflict-Free Replicated Data Types (CRDTs), coining
the term \emph{Merkle-CRDTs} and providing an overview of the different
concepts, properties, advantages and limitations involved. We show how
Merkle-DAGs can act as logical clocks giving Merkle-CRDTs the potential
to greatly simplify the design and implementation of convergent data types
in systems with weak messaging layer
guarantees and a very large number of replicas. Merkle-CRDTs can leverage
highly scalable distributed technologies like DHTs and PubSub algorithms running underneath to take
advantage of the security and de-duplication properties of content-addressing.
Examples of such content-oriented systems could include peer-to-peer content exchange and synchronisation applications
between opportunistically connected mobile devices, IoT devices or user applications running in a web browser.

\end{abstract}

\section{Introduction}

The advent of blockchain technology has generalized the use of peer-to-peer
networking along with cryptographically-directed acyclic graphs, known
as Merkle-DAGs, to implement globally distributed and eventually consistent
data structures in applications such as cryptocurrencies. In these systems,
the Merkle-DAG is a content-addressed data structure used to provide both 
causality information and self-verification of objects that can be easily 
and efficiently shared in trustless peer-to-peer environments. The need to maintain and apply certain 
rules to add new blocks to the blockchains in adversarial scenarios
usually warrants the use of consensus algorithms.

A different approach to obtaining eventual consistency in a distributed system
is by using Conflict-Free Replicated Data Types (CRDTs) \cite{crdts,crdt-study}. CRDTs
are useful in non-adversarial scenarios, where the participating replicas are known
to behave correctly. CRDTs rely on some
properties of the data objects themselves that enable convergence towards a
global, unique state without the need for consensus. CRDTs come in two main
flavours: \emph{state-based CRDTs}\footnote{Also known as \emph{Convergent}
  CRDTs or \emph{CvRDTs}.}---where the states of replicas form a join-semilattice and
are merged under the guarantees afforded by it---and \emph{operation-based
  CRDTs}\footnote{Also known as \emph{Commutative} CRDTs or \emph{CmRDTs}.}
---in which commutative operations are broadcast and applied to the local
state by every replica. Additionally, \emph{$\delta$-CRDTs} are an
optimization of state-based CRDTs to reduce the size of the payloads sent by
the replicas.

Both Merkle-DAGs and CRDTs provide interesting properties: the former allows distributed
systems to take advantage of a content-addressing layer for the resolution/discoverability and self-verification
of data regardless of the source location; the latter allows global state convergence without the
need for---usually complex and expensive---consensus mechanisms. By embedding CRDT objects
inside Merkle-DAG nodes, we obtain the best properties of both worlds, that is, we obtain a convergent
system that can leverage the DAG as a logical clock. This logical clock is provided
and built by every replica without the need for coordination. Replicas can operate undisrupted in loose network environments with
no delivery guarantees. As we will see, a system based on Merkle-CRDTs is fully agnostic to how the system
announces and discovers data among replicas, thus being able to leverage different approaches
like those provided by DHT and PubSub mechanisms without being tied to a particular version of them. This is in stark contrast to traditional consensus algorithms, e.g., Raft, which are tied to particular message dissemination protocols.

We conceive this approach as extremely useful for fully distributed and unstructured peer-to-peer applications,
where the replicas are writers to a common dataset, usually in the form of a database. This is the case for example, in
a distributed and fully replicated file-system, chat group or package repository index. We have found that using the InterPlanetary File System (IPFS) \cite{ipfs} (see Section~\ref{ipfs}) as a 
content-addressed, peer-to-peer decentralised file system and content distribution network, a Merkle-CRDT-based system scales well
to the order of thousands of replicas
which can opportunistically join and depart -- a very common condition when working with mobile, browsers or IoT devices.

The InterPlanetary File System provides a
content-addressed peer-to-peer filesystem \cite{ipfs}, which supports seamless
syncing of Merkle-DAGs with arbitrary formats and payloads, making it a robust
building block for different types of distributed applications like
PeerPad\footnote{PeerPad is a real time p2p collaborative editing tool
  (\url{https://peerpad.net}).} or OrbitDB\footnote{OrbitDB is a
  peer-to-peer database for the decentralized web
  (\url{https://github.com/orbitdb/orbit-db}).}, both powered by CRDTs and
IPFS.
IPFS not only delivers the right environment for working with content addressed data (like the Merkle-DAG that we will use), 
but also allows our CRDT application to be fully detached from the lower levels
of the distributed system: networking transports, discovery and data transfer facilities come in modules
which can be swapped and tweaked independently. Leveraging such system greatly simplifies the design
and optimization of the CRDT layer so that it adapts well to the use case it is meant to serve.

In this paper we formalize what we refer to as \emph{Merkle-CRDTs}. The
goal is to provide an overview of their properties, advantages and
limitations, so that it can set the ground layer for future research and
optimizations in the space. For example, Merkle-CRDTs allow  
building fully distributed key-value stores in real-world systems in networks with no message delivery
guarantees and fully flexible replica sets that can grow and shrink at any time 
without impacting the CRDT layer.

The contributions of this paper are as follows:
\begin{itemize}
\item We define \emph{Merkle-Clocks} as Merke-DAG-based logical clocks,
to represent causality information in a distributed system. Embedding
causality information using Merkle-DAGs is at the core of cryptocurrencies and
source control systems like Git but they are rarely considered separately as
a type of logical clock. We demonstrate that Merkle-Clocks can be used in
place of other logical clocks traditionally used by CRDTs like version vectors
and vector clocks. We show that Merkle-Clocks can, in fact, be seen as CRDT
objects themselves, which can be synced and merged and for which we can formally prove
eventual consistency across different replicas.
\item We define \emph{Merkle-CRDTs}
as a general purpose transport and persistence layer for CRDT payloads which
leverages the properties of Merkle-Clocks, using the DAG-Syncer and the
Broadcaster to provide per-object causal consistency by design. This
enables the use of simple CRDT types in systems with weak messaging layer
guarantees and large number of replicas.
\end{itemize}

Our intention with this paper is to show that eventual consistency between replicas can be achieved \emph{independently of the underlying transport mechanisms and in a setting where all peers are equal}. This means that Merkle-CRDTs can be implemented on top of any underlying peer discovery and routing system (e.g., DHT, PubSub). We argue that this is a very powerful concept that has the potential to lead to new eventual consistency system designs. It is also fundamentally different to traditional consensus algorithms, such as Raft, where at any given time there needs to be a leader peer that collects the latest state from all other peers. As such, we do not provide an evaluation against these traditional approaches as their requirements and design principles are fundamentally different.

The rest of the paper is organised as follows:
In Section~\ref{background}, we start by introducing relevant background concepts and prior art. In Section~\ref{system-model}, we expose the characteristics of our system model and introduce
the facilities needed to store and sync Merkle-CRDTs.

In Section~\ref{merkle-clocks}, we introduce Merkle-Clocks and, building on the previous sections, in Section~\ref{merkle-crdts}, we define Merkle-CRDTs. We discuss how different CRDT
payloads (whether operation-based, state-based or $\delta$-based) benefit from
Merkle-CRDTs. Finally, we describe some of the limitations and inefficiencies
of Merkle-CRDTs and introduce techniques to overcome them in Section~\ref{limits-optimisations}.

\section{Background \& Related Work}\label{background}

\subsection{Eventual consistency}

\textit{Eventual Consistency} (EC) in distributed systems refers to the situation where the state may not be the same across replicas of the system but, given enough time and perhaps after network
partitions, downtime and other eventualities have been resolved, the system
design will ensure that the state becomes the same everywhere.

The main weakness of the eventual consistency definition is that it offers no
guarantees as to when the shared state will converge or how much the
individual states will be allowed to diverge until then\footnote{EC only
  provides a \emph{liveness} guarantee: the system will not become stuck when making
  progress to converge.}.  \emph{Strong eventual consistency} (SEC) addresses
these issues by establishing an additional safety guarantee: if two replicas
have received the same updates, their state will be the same.

Consensus algorithms or, more important to this paper, Conflict-Free
Replicated Data Types (CRDTs) are ways to achieve (strong) eventual
consistency in a distributed system.

\subsection{Merkle DAGs}

A \emph{Directed Acyclic Graph (DAG)} is a type of graph in which edges have
direction and cycles are not allowed.
For example, a linked list like $A
\rightarrow B \rightarrow C$ is an instance of a DAG where $A$ references
$B$ and so on. We say that $B$ is \emph{a child} or
  \emph{a descendant of} $A$, and that \emph{node $A$ has a link to $B$}.
Conversely $A$ is a \emph{parent of} $B$. We call nodes\footnote{Throughout the paper, we use the term \emph{replica} to refer to the physical machine of a network node and \emph{node} to refer to bundled content addressed by a single \emph{identifier}.} that are not children to any
other node in the DAG as the \emph{root nodes}.
A Merkle-DAG is a DAG where each node has an identifier and this is the result
of hashing the node's contents ---any opaque payload carried by the node and
the list of identifiers of its children--- using a cryptographic hash function
like SHA256.
This brings some important considerations:
\begin{enumerate}[label=\alph*)]
\item Merkle-DAGs can only be constructed from the leaves, that is, from nodes
  without children. Parents are added after children because the children's
  identifiers must be computed in advance to be able to link them.
\item every node in a Merkle-DAG is the root of a (sub)Merkle-DAG itself, and
  this subgraph is \emph{contained} in the parent DAG\footnote{Merkle-DAGs are
    similar to Merkle Trees \cite{merkle-trees} but there are no balance
    requirements and every node can carry a payload. In DAGs, several branches
    can re-converge or, in other words, a node can have several parents, or be part of several Merkle DAGs.}.
\item Merkle-DAG nodes are \emph{immutable}. Any change in a node would alter
  its identifier and thus affect all the ascendants in the DAG, essentially
  creating a different DAG.
\end{enumerate}
Identifying a data object (like a Merkle-DAG node) by the value of its hash is
referred to as \emph{content addressing}. Thus, we name the node identifier as \emph{Content Identifier} or CID.


For example, in the previous linked list, assuming that the payload of each node
is just the CID of its descendant would be: $A=Hash(B) \rightarrow B=Hash(C)
\rightarrow C=Hash(\emptyset)$. The properties of the hash function ensure
that no cycles can exist when creating Merkle-DAGs.
\footnote{Hash functions
  are one way functions. Creating a cycle should then be impossibly difficult,
  unless some weakness is discovered and exploited.}.

Merkle-DAGs are self-verified and immutable structures. The CID of a node is
univocally linked to the contents of its payload and those of all its
descendants. Thus two nodes with the same CID univocally represent exactly the
same DAG. This will be a key property to efficiently sync Merkle-CRDTs without
having to copy the full DAG, as exploited by systems like IPFS.

Merkle-DAGs are widely used. Source control systems \cite{current-concepts-in-version-control-systems} like Git \cite{git} use them to
efficiently store the repository history in a way that enables
de-duplicating the objects and detecting conflicts between branches. In distributed databases like Dynamo \cite{dynamo}, Merkle-Trees are used for
efficient comparison and reconciliation of the state between replicas. In Hash
Histories \cite{hash-histories}, content addressing is used to refer to a
Merkle-Tree representing a state.


Merkle-DAGs are also the foundational block of blockchains---they can be seen
as a Merkle-DAG with a single branch---and their most common application:
cryptocurrencies (e.g., \cite{iota}, \cite{dagcoin},
\cite{byteball}). Cryptocurrencies like Bitcoin \cite{bitcoin} benefit from
the embedded causality information encoded in the chain: transactions in a
block deeper in the chain always happened before those of earlier blocks.  One
of the main issues in cryptocurrencies is to make all participating peers
agree about the tip/head/root of the chain.  Among other things, the
non-commutative nature of some transactions, requires a consensus mechanism
which enforces that only valid blocks become the new roots.


One commonality in many of these systems is that the Merkle-DAG implicitly
embeds causality information.  The DAG can show that a certain transaction
precedes another or that a Git commit needs to be merged rather than
fast-forwarded. This will be one of the properties that we use in Merkle-CRDTs
and that this paper makes explicit and puts in contrast with other
causality-encoding mechanisms known as \emph{logical clocks}.

\subsection{Logical clocks}

The design of causally-convergent systems involves the reconciliation of
diverging state versions among different replicas when, for example, events
occur concurrently.

This requires that we are able to identify whether two
events actually happened concurrently and whether two states are actually
different because of concurrent updates or other reasons, such as one replica
having received more updates.

The problem is, essentially, tracking the order in which different
events happened. For example, given multiple writes of a value to a register in
different replicas, we would expect the final value in the registry to be that
of the \emph{last} write.

Ideally, we should be able to order all the events in the system\footnote{This
  means establishing a \emph{total strict order} for all the events.} so that
we can identify which was the actual \emph{last} update to the register.

Tagging events with timestamps can give us this information: if all events are
timestamped, any replica may establish the order in which they happened and
use that information to decide what the final state should look like. However,
In distributed systems, it is not possible to use timestamps reliably
\cite{Neville}, as not every replica can be perfectly synced to a global
time. ``Wall clocks'' can also easily be simulated or spoofed, which is
problematic in peer-to-peer systems with no trust involved.

\emph{Logical clocks} are the alternative to global time. They provide ways to
encode causal information between events known to different actors in a
distributed system.

The basic idea is that, although we may not know the order in which all events
happened globally, every replica knows at least the order of events issued by itself.
Any other replica that receives that information will then know that any
events later issued by itself come after those.  This is, in essence, what is
known as \emph{causal history}.

Logical clocks are representations of causal histories
\cite{why-logical-clocks-are-easy} and provide a partial ordering
between events. That is, given two events $a$ and $b$, logical clocks should
be able to tell us if $a$ \emph{happened before} $b$ ($a \rightarrow b$), or
vice-versa ($b \rightarrow a$), or if both $a$ and $b$ happened concurrently
($a \parallel b$).

%




The practical implementation of logical clocks usually involves metadata which
travels attached to every event in the system. One of the most common forms of
logical clocks are version vectors \cite{version-vectors}: every
replica maintains and broadcasts a vector that tracks on which version the
state of all the replicas is. When a replica performs a
modification of the state, it increases its version. When a replica merges a
state from a different replica, it takes the highest between the local
versions and the versions provided by the other replica along with the event.
Thus, given two events $a$, $b$, with version vectors $\mathcal{V}^{a}$,
$\mathcal{V}^{b}$: $a \rightarrow b$ if $\mathcal{V}_i^{a} \leq
\mathcal{V}_i^{b}$ for each position $i$ in the vectors.  If $a
\not\rightarrow b$ and $b \not\rightarrow a$, by that definition, $a$ and $b$
are concurrent.

As we see, version vectors are compact because they do not need to store the
full causal history but merely a number indicating how long the history is for
every replica.  Version vectors depend on the number of replicas, so they may
need further optimizations to work well in scenarios with many replicas or
where the number of replicas is not stable.

In this paper we demonstrate that a Merkle-DAG can act as a logical clock. \emph{Merkle-Clocks}, as we will
show, provide a different set of properties but encode the same causal information about events.

\subsection{Conflict-Free Replicated Data Types (CRDTs)}

CRDTs are data types which provide strong eventual consistency among different
replicas in a distributed system by requiring certain properties from the
state and/or the operations that modify it. Additionally, CRDTs also feature
monotonicity. The concept of monotonicity applied to data types is the notion
that every update is an inflation, making the state grow, not in size, but in
respect to a previous state. This implies that there will always be an order
between states. Monotonicity implies that rollbacks on the state are not
necessary regardless of the order in which updates happen.

There are two prominent types of CRDTs: \emph{state-based} and \emph{operation-based}
CRDTs.  In state-based CRDTs, all the states in the system ---that is, the
states in different replicas and different moments--- form a monotonic
join-semilattice. That means that, for any two states $X$ and $Y$, both can be
\emph{"joined"} (\emph{merged}, or form a \emph{union})
($\sqcup$) and the result is a new state corresponding to the
Least-Upper-Bound (LUB) of the two \cite{crdts}. In other words, every
modification made to a state by a replica must be an inflation and the union
of two states $X$ and $Y$ is the minimal state capable of containing both $X$
and $Y$ and not more (the LUB). A join-semilattice is thus a partially ordered
set and its LUB is the smallest state capable of \emph{containing}
all the states in the semilattice. This implies that the $\sqcup$ operation
must be idempotent ($X \sqcup X = X$), commutative ($X \sqcup Y = Y \sqcup X$)
and associative ($(X \sqcup Y) \sqcup Z = X \sqcup (Y \sqcup Z)$).

Replicas in a state-based CRDT modify their state---or inflate it---and
broadcast the resulting state to the rest of replicas\footnote{An important
  note here is that CRDTs are just data types. The transmission policy of CRDT
  objects between replicas is independent. Some CRDTs are, by design,
  better suited to some broadcasting mechanisms than others and can facilitate
  optimizations such as broadcasting only to a random subset rather than to
  every replica.}. Upon receiving the state, the other replicas merge
it with the local state. The properties of the state ensure that, if the replicas
have correctly received the states sent by other replicas---and
vice-versa---, they will eventually converge.

Operation-based CRDTs \cite{crdts}, on the other side, do not enforce any
property on the state itself but on the operations used to modify it, which 
must be commutative (at least with regard to a different
  operation issued at the same time (concurrently). The replicas
broadcast the operations and not the states. If two operations happen at the
same time in two replicas, the order in which other
replicas apply them does not matter: the resulting states will be the same.

It follows that, if an operation broadcast does not arrive to a replica ---for
example due to a network failure---, that replica will never be able to apply
it and the states will not converge.  Thus, unlike state-based CRDTs, eventual
consistency in operation-based CRDTs requires a reliable messaging layer that
eventually delivers all operations \cite{making-op-based-op-based}. Additional
constraints may be necessary, for example, if operations are not idempotent:
in that case, the messaging layer should ensure that each operation is
delivered exactly once.


Some operation-based CRDTs may also require causal
delivery: if a replica sends operation $a$ before $b$ ($a \rightarrow b$),
then $a$ should always be delivered before $b$ to a different replica.

These properties and requirements in both state and operation-based CRDTs
ensure \emph{per-object causal consistency}: updates to a state will maintain
the causal relations between them. For example, in a Grow-Only Set (G-Set),
when a replica adds element $A$ and then element $B$, no other replica will
ever have a set where $B$ is part of the set but $A$ is not.
\footnote{This is
  clear for an operation-based implementation of a G-Set (assuming causal
  delivery of the operations).  The state-based implementation of a G-Set
  involves sending the full set.  Thus, the event adding $B$ is a set which
  already contains $A$: there will not be a set where $B$ is present but
  not $A$, even if the event that added $A$ was lost or arrives later.}.

Logical clocks, as seen in the previous section, are commonly used to
implement CRDT types: they are useful to identify when two updates happen
concurrently and need merging. CRDTs have been successfully used and optimized in different applications and
distributed databases, Basho's Riak \cite{riak-dt-map,riak-bigger} being one
of the most prominent examples.

\subsection{Sync Protocols in Information-Centric Networks}

There are multiple types of logical
clocks that are similar to version vectors discussed earlier, but fulfil different needs or
address some of their shortcomings: vector clocks \cite{vector-clocks},
bounded version vectors \cite{bounded-version-vectors}, dotted version vectors
\cite{dotted-version-vectors}, tree clocks \cite{tree-clocks} or interval tree
clocks \cite{interval-tree-clocks} are some of them.

There has been a recent body of work in distributed dataset synchronisation in the area of Information-Centric Networks (ICN) \cite{tr-shang2017sync-survey}, \cite{vectorsync}, \cite{chronosync}, \cite{chronoshare}. Information-Centric architectures are advocating direct content naming at the network layer and subsequent routing and forwarding (by core network routers), based on content names and (in some cases) longest-prefix matching.

ChronoSync \cite{chronosync} is utilising the features of the Named-Data Networking architecture \cite{zhang2014named} to synchronise state between different datasets. ChronoSync is a data-layer mechanism, which, however, takes advantage of the hierarchical and flexible naming scheme that NDN is building on. Inspired by Merkle Trees, ChronoSync is using cryptographic digests and filters to synchronise datasets between peers. ChronoSync takes advantage of the name-based nature of the underlying network (NDN) and is assigning a unique publishing prefix to each peer. This unique prefix, together with sequence numbers and network layer persisting/long-lived ``Interest packets" is replacing much of what other approaches attempt to do with clocks and CRDTs. While integrating sync functionality at the network layer allows for more native designs, some features are inevitably lost. For example, ChronoSync cannot deal with simultaneous (concurrent) data publication. RoundSync \cite{tr-shang2017sync-survey} is partially solving this problem by splitting the sycnhronisation process in rounds.

VectorSync \cite{vectorsync} is an enhanced version of ChronoSync \cite{chronosync} which uses version vectors to make synchronisation between peers more efficient. Version vectors are more efficient in detection of inconsistencies, than simple message digests, as mentioned earlier. However, similarly to other proposals in the area, VectorSync needs to realise a `leader-based membership management' in order to deal with active members that update the state of the dataset.

The integration of distributed dataset synchronisation features natively at the network layer of the network is clearly an advanced endeavour, which comes with its own challenges. We believe that the advantage of ``transport-agnostic" state synchronisation brought by Merkle-CRDTs can apply and improve the performance of protocols such as ChronoSync, RoundSync or VectorSync. On the other hand, handling Merkle-CRDT-based state synchronisation directly through named network objects brings standard ICN advantages to Merkle-CRDTs. As such we consider those two distinct approaches to state synchronisation as complementary.

\subsection{IPFS: The InterPlanetary File System}\label{ipfs}

IPFS~\cite{ipfs} is a content-addressed, distributed filesystem. 
It uses a Distributed Hash Table (DHT) to announce and discover
which replicas (or peers) provide certain Merkle-DAG nodes. It implements a
node-exchange protocol called \emph{bitswap} to retrieve DAG nodes from any
provider. IPFS is built on top of libp2p \cite{libp2p},
a modular network protocol stack for P2P networks, which additionally provides 
 efficient broadcasting mechanisms primarily based on publish-subscribe models \cite{gossipsub}.

IPFS also uses IPLD, the \emph{InterPlanetary Linked Data Format} \cite{ipld},
a framework to describe Merkle-DAGs with arbitrary node formats and support
for multiple types of CIDs \cite{multiformats}, making it very easy to create
and sync custom DAG nodes.

These features make IPFS a suitable layer on which to implement Merkle-CRDTs, as it
provides the necessary mechanisms to discover, route and announce content in
potentially very large networks. This is not to say that other transport mechanisms are not suitable to build Merkle-CRDTs on top.



\section{System model \& Assumptions}\label{system-model}

Our Merkle-CRDT approach is intended to be both simple and facilitate the use
of CRDTs in peer-to-peer distributed systems with large number of replicas and no message delivery
guarantees (i.e., unreliable transports).

We assume the presence of an asynchronous messaging layer which provides a
communication channel between separate replicas. This channel is managed by
two facilities which every replica exploits: the \emph{DAG-Syncer} and the
\emph{Broadcaster} components (defined below).

We assume that messages can be dropped, reordered, corrupted or duplicated. It is not
necessary to know beforehand the number of replicas participating in the
system. Replicas can join and leave at will, without informing any other
replica. There can be network partitions but they are
resolved as soon as connectivity is re-established and a replica broadcasts a
new event.

Replicas may have durable storage, depending on their own requirements and
data types. Using Merkle-CRDTs, new replicas and crashed replicas without durable 
storage will be able to eventually re-construct the complete state of the 
system as long as at least one other replica is in the latest system state.

\subsection{The DAG-Syncer component}

A \emph{DAG-Syncer} is a component that enables a replica to obtain remote
Merkle-DAG nodes from other replicas given their content identifiers (CIDs)
and to make its own nodes available to other replicas.  Since a node contains links
to their direct descendants, given the root node's CID, the DAG-Syncer
component can be used to fetch the full DAG by following the
links to children in each node. Thus, we can define the DAG-Syncer as follows:

\begin{definition}{(DAG-Syncer).}
A DAG-Syncer is a component with two methods:
\vspace{-0.2cm}
\begin{itemize}[noitemsep]
\item \texttt{Get(CID) : Node}
\item \texttt{Put(Node)}
\end{itemize}
\end{definition}
\vspace{-0.2cm}
We do not specify any more details such as how the protocol to announce and
retrieve nodes looks like. Ideally, the DAG-Syncer layer should not impose any
additional constraints on the system model.  Our approach relies on the
properties of the DAG-Syncer and Merkle-DAGs to tolerate all the network
contingencies described above.

\subsection{The Broadcaster component}

A \emph{Broadcaster} is a component to distribute arbitrary data from one
replica to all others (directly or through relays).  Ideally, the payload will
reach every replica in the system, but this is not a requirement for every
broadcast message:

\begin{definition}{(Broadcaster).}
A Broadcaster is a component with one method:
\vspace{-0.2cm}
\begin{itemize}[noitemsep]
\item \texttt{Broadcast(Data)}
\end{itemize}
\end{definition}
\vspace{-0.2cm}
\subsection{IPFS as a DAG-Syncer and Broadcaster component}

The components above can be realised by using
IPFS (as introduced in Section \ref{background}).
IPFS can act as the DAG-Syncer, while one of the PubSub mechanisms provided by 
its libp2p layer can perform the tasks of the Broadcaster component.

Such an implementation should allow extreme scalability of the replica set in general. The peers in the
network do not need to be fully connected to everyone else and the system is extremely modular
and configurable to fit both small devices and large storage servers. The choice of settings
and implementations will affect the performance of the system under different circumstances
and network topologies but is independent from the Merkle-CRDT objects and datatype.


\section{Merkle-Clocks}\label{merkle-clocks}

\subsection{Overview}

A Merkle-Clock $\mathscr{M}$ is a Merkle-DAG where each node represents an
event. In other words, given an event in the system, we can find a node in
this DAG that represents it and that allows us to compare it to other events.

The DAG is built by merging other DAGs (those in other replicas) according to
some simple rules. New events are added as new root nodes (parents to the
existing ones). Note that the Merkle-Clock may have several roots at a given time.

For example, given $\mathscr{M}_\alpha$ and $\mathscr{M}_\beta$ ($\alpha$ and
$\beta$ being the single root CIDs in those DAGs\footnote{In the example we
  assume, without loss of generality, that we start with DAGs containing a
  single root instead of several.}):

\begin{enumerate}
\item If $\alpha = \beta$ no action is needed, as they are the same DAG.
\item else if $\alpha \in \mathscr{M}_\beta$, we keep $\mathscr{M}_\beta$ as
  our new Clock, since the history in $\mathscr{M}_\alpha$ is part of it
  already. We say that $\mathscr{M}_\alpha < \mathscr{M}_\beta $ in this case.
\item else if $\beta \in \mathscr{M}_\alpha$, we keep $\mathscr{M}_\alpha$ for
  the same reason.  We say that $\mathscr{M}_\beta < \mathscr{M}_\alpha $
  in this case.
\item else, we \emph{merge} both Clocks by keeping both DAGs as they are and
  thus having two root nodes, those referenced by $\alpha$ and $\beta$.
  Note that $\mathscr{M}_\alpha$ and
  $\mathscr{M}_\beta$ could be fully disjoint or not, depending on whether
  they share some of their deeper nodes. If we wish to record a new event, we
  can create a new root $\gamma$ with two children, $\alpha$ and
  $\beta$.
\end{enumerate}

We can already see that, by determining if one Merkle-Clock is included in another,
we are introducing the notion of order among Merkle-Clocks. In the same way, we have a notion of
order among the nodes in each clock, since events that happened earlier will
always be descendants of events that happened later. Additionally, we have
introduced a way to merge Merkle-Clocks according to how they compare. The
resulting Merkle-Clock always includes the causality information from both
Merkle-Clocks. This eventually means that the causality information stored in
Merkle-Clocks in every replica will converge to the same Merkle-Clock after
merging.


The causal order provided by Merkle-Clocks is embedded when building
Merkle-DAGs with similar rules and usually overlooked as something very
intuitive. It is important, however, to formalize how we define order
between Merkle-Clocks and to prove that the causality information is
maintained when they are synced and merged. This is the subject of the next section and will be an important property for Merkle-CRDTs.

\subsection{Merkle-Clocks as a convergent, replicated data type}

This section formalizes the definition of Merkle-Clocks and their
representation as Merkle-Clock DAGs. We will show that Merkle-Clock DAGs can
be seen as a Growing-Set (G-Set) CRDT and therefore converge in multiple
replicas.

%

Let $\mathscr{S}$ be the set of all system events:

\begin{definition}{(Merkle-Clock Node).} A Merkle-Clock Node $n_\alpha$ is a triple:
\[ (\alpha, e_\alpha, \mathcal{C}_\alpha) \]
which represents an event $e_\alpha \in \mathscr{S} $, with $\alpha$ being the
node CID and $\mathcal{C}_\alpha$ being the CID-set of $n_\alpha$'s direct
descendants.
\end{definition}

\begin{definition}{(Merkle-Clock DAG).} A Merkle-Clock DAG is a pair:
\[ \langle \mathbb{N}, \leq \rangle \]
where $\mathbb{N}$ is a set of immutable DAG-nodes and a partial order $\leq$ on
$\mathbb{N}$, defined as follows:
\[ n_\alpha, n_\beta \in  \mathbb{N}: n_\alpha < n_\beta \Leftrightarrow n_\alpha \text{ is a descendant of } n_\beta \]

In other words, $n_\alpha < n_\beta$ if there is a path of linked nodes which
goes from $n_\beta$ to $n_\alpha$.
\end{definition}

In order to maintain this relationship, the Merkle-Clock DAG must be built
with the following Implementation Rule:

\begin{ir}
Every new event in the system must be represented as a new root node to the
existing Merkle-Clock DAG(s). In particular, the $\mathcal{C}$ set must 
contain the CIDs of the previous roots.
\end{ir}

\begin{definition}{(Merkle-Clock).}
A Merkle-Clock ($\mathscr{M}$) is a function which given an event $e_\alpha
\in \mathscr{S}$ returns a node from the Merkle-Clock DAG $\mathbb{N}$:
\[ \mathscr{M} \colon \mathscr{S} \rightarrow \mathbb{N}  \]

\end{definition}

\begin{remark}
A Merkle-Clock satisfies the Strong Clock condition
\cite{Lamport-clocks}. We see that every node represents a later event than
that of its children:
\[ \forall (\beta, e_\beta, \mathcal{C}_\beta) \in \mathbb{N}:
\forall \alpha \in \mathcal{C}_\beta: e_\alpha \rightarrow e_\beta \]
Since every event is the root of a (sub)DAG built using the implementation
rule, we can immediately see that earlier Merkle-Clock values are descendants
of the later ones:
\[ \mathscr{M}(e_\alpha) < \mathscr{M}(e_\beta)
\Leftrightarrow e_\alpha \rightarrow e_\beta \]
\end{remark}
We can now define a join-semilattice of Merkle-Clocks DAGs as a pair:
\[ \langle \mathbb{J}, \subseteq_\mathbb{J} \rangle \]
where $\mathbb{J}$ is a set of Merkle-Clocks DAGs and $\subseteq_\mathbb{J}$ a
partial order over that set defined as follows. Given $\mathbb{M},\mathbb{N}
\in \mathbb{J}$:
\[ \mathbb{M} \subset_\mathbb{J} \mathbb{N} \Leftrightarrow \forall m \in \mathbb{M}, \exists n \in \mathbb{N} \mid m < n \Leftrightarrow \mathbb{M} \subset \mathbb{N} \]
Note that $m < n$, means that $m$ is a descendant of $n$ and thus must belong
to the same DAG, then $\subset_\mathbb{J}$ simply means that $\mathbb{M}$ is a
subset of $\mathbb{N}$.

This allows us to define the Least-Upper-Bound of two Merkle-Clocks DAGs
($\sqcup_\mathbb{J}$) as the regular union of the sets:
\[ \mathbb{M} \sqcup_\mathbb{J} \mathbb{N} =  \mathbb{M} \cup \mathbb{N} \]

Unsurprisingly, the Merkle-Clock representation corresponds in fact to a
Grow-Only-Set (G-Set) in the state-based CRDT form \cite{crdt-study}.  The
elements of the set are immutable, cryptographically linked and represent the
events in the system. When the DAGs are disjoint, the resulting DAG will
include the roots from both $\mathbb{N}$ and $\mathbb{M}$. That is the
equivalent of having several events without causal relationship. Causality
information about DAG-merge events can be optionally included after the union
of the DAGs by creating a new unique root following the implementation rule.

In the next section we will see how the properties of Merkle-DAGs allow
syncing Merkle-Clocks in a more efficient manner than regular state-based
G-Sets.

\subsection{The Merkle in the Clocks: properties of Merkle-Clocks }

We have so far defined a way to encode causality information per replica and
ensured that two replicas can merge their Merkle-Clocks. Now we will see
how the properties of Merkle-DAGs allow the use of a \emph{pull} rather than a
\emph{push} approach which, together with content-addressing, enables
efficient clock sync between replicas and overcomes the effect of network partitions or contingencies. 
The steps to Merkle-Clock synchronisation between replicas are given below.

\begin{enumerate}
\item Broadcasting the Merkle-Clock requires broadcasting only the current
  root CID.  The whole Clock is unambiguously identified by the CID of its
  root and its full DAG can be walked down from it as needed.
\item The immutable nature of a Merkle-DAG allows every other replica to perform quick
  comparisons and pull/fetch only those nodes that it does not already have.
\item Merkle-DAG nodes are self-verified, through their CID, and, therefore, immune to corruption and
  tampering. Hence, they can be fetched (pulled) from any source willing to provide them, trusted or not.
\item Identical nodes are de-duplicated by design: there can only be one unique
 representation for every event.
\end{enumerate}

In practice, every replica just fetches the \emph{delta} causal histories from
other replicas without the need to build those deltas explicitly anywhere in the
system. A completely new replica with no previous history will fetch the full
history automatically\footnote{This is how peers participating in
  cryptocurrency minning sync their ledgers.}.

Merkle-Clocks can replace version clocks and other logical clocks that are
usually part of CRDTs.  This comes with some considerations:

\begin{itemize}
\item By using Merkle-Clocks we can decouple the causality information from the number
  of replicas, which is a common limitation in version clocks. This makes it
  possible to reduce the size of the messages when implementing CRDTs and,
  most importantly, solves the problem of keeping clocks working when
  replicas randomly join and leave the system.
\item On the downside, the causal information grows with every event and
  replicas store potentially large histories even if the event information is
  consolidated into smaller objects.
\item Keeping the whole causal history enables new replicas to sync events
  from scratch out-of-the-box, without having to explicitly send system
  snapshots to newcomers. However, that syncing may be slow if the history is
  very large. We will explore, along with Merkle-CRDTs, potential
  optimizations in this regard.
\end{itemize}

A significant advantage of Merkle-Clocks over traditional version clocks is that they can also deal with several types of network anomalies:

\begin{itemize}
\item \emph{Dropped messages} may prevent  other replicas from learning about new roots.
  But since every Merkle-Clock DAG is superseeded by future DAGs and every
  download fetches all the missing parts of a DAG, network partitions and
  replica downtimes do not have an effect on the overall system and will begin
  to heal automatically once the issues are resolved.
\item \emph{Out of order delivery} poses no problem for the same reasons. The
  missing DAG will be fetched and processed in order.
\item \emph{Duplicated messages} are just ignored by replicas as they are already
  incorporated into their Merkle-Clocks.
\item \emph{Corrupt messages} come in two forms:
\begin{enumerate*}[label=\alph*)]
\item if the message broadcasting a new root is corrupted, then it will be a
  hash corresponding to a non-existent DAG that cannot be fetched by the
  DAG-Syncer and will be eventually ignored;
\item if a DAG node is corrupted on download, the DAG-Syncer component (or the
  application) can discard it if its CID does not match the downloaded
  content.
\end{enumerate*}
\end{itemize}

As we showed in the previous section, Merkle-Clocks represent a strict partial
order of events. Not all events in the system can be compared and
ordered. For example, when having multiple roots, the Merkle-Clock cannot say
which of the events happened first.

A total order can be useful \cite{Lamport-clocks} and could be obtained, for example, 
by considering concurrent events to be equal. Similarly, a strict total order 
could be built by sorting concurrent events by the CID of their nodes or by any other
arbitrary user-defined strategy based on additional information attached 
to the clock nodes. Any such approach would qualify as \emph{data-layer conflict resolution}.


\section{Merkle-CRDTs: Merkle-Clocks with payload}\label{merkle-crdts}

\begin{definition}{(Merkle-CRDT).}
A Merkle-CRDT is a Merkle-Clock whose nodes carry an arbitrary CRDT payload.
\end{definition}

Merkle-CRDTs keep all the properties seen before for Merkle-Clocks. However,
for the payloads to converge, they need to be convergent data types (CRDTs)
themselves. The advantage is that Merkle-Clocks already embed
ordering and causality information which would otherwise need to travel embedded
in the CRDT objects (usually in the form of other logical clocks).
or be provided by a reliable messaging layer.

Thus, the implementation of a Merkle-CRDT node is:
\[ (\alpha, \mathit{P}, \mathcal{C}) \]
with $\alpha$ being the content identifier, $\mathit{P}$ an opaque data
object with CRDT properties and $\mathcal{C}$ the set of children
identifiers.


\subsection{Per-object causal consistency and gap detection}

The directed-link nature of Merkle-CRDTs, which allows traversing the full
causal history of the system in the order of events, provides all the necessary
properties to ensure per-object \emph{causal consistency} and \emph{gap
  detection} by design without modifying our system model.

This means that Merkle-CRDTs are very well suited to carry operation-based
CRDTs as they can ensure that no operation is lost or applied in
disorder\footnote{Recall that the Merkle-Clock provides a strict partial
order of events. In this case, two non-concurrent operations applied to
an object will be sortable by the clock.}.

To facilitate the task of processing CRDT payloads in Merkle-CRDTs, in the next
section we present a general and simple (non-optimized) anti-entropy algorithm
that can be used to obtain per-object causal consistency for any CRDT embedded 
object.

\subsection{General anti-entropy algorithm for Merkle-CRDTs}

\begin{definition}{(General anti-entropy algorithm for Merkle-CRDTs).}

Let $\mathcal{R}^A$ and $\mathcal{R}^B$ be two replicas using Merkle-CRDTs
with $\mathscr{M}_\alpha$ and $\mathscr{M}_\theta$
respectively as their
current Merkle-CRDT DAG.

\begin{enumerate}

\item $\mathcal{R}^B$ issues a new payload by creating a new DAG node $(\beta,
  P, \lbrace \theta \rbrace)$ and adding it as the new root to its
  Merkle-CRDT, which becomes $\mathscr{M}_\beta$.

\item $\mathcal{R}^B$ broadcasts $\beta$ to the rest of replicas in the
  system.

\item $\mathcal{R}^A$ receives the broadcast of $\beta$ and retrieves the full
  $\mathscr{M}_\beta$. It does this by starting from the root $\beta$ and
  walking down the DAG using the DAG-Syncer component to fetch all the nodes
  that are not in $\mathscr{M}_\alpha$, while collecting their CIDs in a
  CID-Set $\mathcal{D}$. Given the inherent properties of DAGs, for any CID already in $\mathscr{M}_\alpha$ the
  whole sub-DAG can be skipped.

\item If $\mathcal{D}$ is empty, no further action is
  required. $\mathcal{R}^A$ must have already processed all the payloads in
  $\mathscr{M}_\beta$. This means that $\mathscr{M}_\beta \subseteq
  \mathscr{M}_\alpha$.

\item If $\mathcal{D}$ is \emph{not} empty, we sort the CIDs in $\mathcal{D}$
  using the order provided by the Merkle-Clock.

  %
 
  We can skip the ordering if causal delivery is not a requirement in
  our system.  The amount of items in $\mathcal{D}$ will depend on the amount
  of concurrency in the system and how long the two Merkle-CRDTs have been
  allowed to diverge, but should be small under normal circumstances.

\item $\mathcal{R}^A$ processes the payloads associated with the nodes
  corresponding to the CIDs in $\mathcal{D}$, from the lowest to the highest.

\item If $\alpha \in \mathcal{D}$, then $\mathscr{M}_\alpha \subseteq
  \mathscr{M}_\beta$ and $\mathscr{M}_\beta$ becomes the new local Merkle-CRDT
  in $\mathcal{R}^A$.

\item else, $\mathscr{M}_\alpha \not\subset \mathscr{M}_\beta$ and
  $\mathscr{M}_\beta \not\subset \mathscr{M}_\alpha$.  $\mathcal{R}^A$ keeps
  both nodes as roots.

\end{enumerate}

\end{definition}


\subsection{Operation-based Merkle-CDRTs}

\begin{definition}{Operation-based Merkle-CRDTs are those in which nodes embed
an operation-based CRDT payload}.
\end{definition}

Operation-based Merkle-CRDTs are the most natural application of Merkle-CRDTs.
Operations are easy to define, as they just need to be commutative, so that the resulting
state will be the same in every replica regardless of the order in which they
have received the operations. However, that also means that for states to converge,
every operation must be received. A reliable messaging layer
\cite{making-op-based-op-based} is then a prerequisite for convergence, but in real
world networks with a large number of replicas it is usually not possible to ensure
that no message is lost. This leaves us with complex workarounds, like additional
causality payloads, buffering and retry mechanisms that must accompany the 
CRDT implementation, turning what should be a simple CRDT implementation into
something considerably more complicated.

Merkle-DAGs provide all the properties of a messaging layer where messages are
always delivered in order, verified and never repeated nor dropped.  Thus,
Merkle-CRDTs enable operation-based CRDTs in contexts where they could not be
easily used before.

As we saw, thanks to the Merkle-DAG in which they are embedded, each replica
only needs the missing parts of the DAG and these can be fetched once the root
is known. This includes new replicas joining the
system, which will be able to fetch and apply all operations. We do not need
to keep knowledge of the full replica set and place the responsibility of
efficient broadcast in the \emph{Broadcaster} component.

%

\subsection{State-based Merkle-CRDTs}

\begin{definition}{State-based Merkle-CRDTs are those in which nodes embed
a state-based CRDT payload}.
\end{definition}

Embedding full states in each Merkle-CRDT node is counter-intuitive since 
state-based CRDTs already provide per-object causal consistency and
can cope with unreliable message layers by design.

Moreover, although the final state would result from the merge of all the
states in the Merkle-CRDT nodes, the \emph{DAG-Syncer} component would still
need to store those states, something prohibitive when working
with large state objects. That said, Merkle-CRDTs remove the need to
attach causality metadata and detach it from the number of replicas,
which might be of interest for state-based CRDTs with very small states
in comparison to the number of replicas.

A more interesting approach is that of $\delta$-CRDTs \cite{delta-crdts}
which, instead of broadcasting full states, are able to send smaller
sections (deltas). $\delta$-mutations, as these objects are called, can be
merged downstream just like any full state would be, without the need for
changing the semantics of the \emph{union} operation.  It follows that
multiple deltas can be merged to form what is known as $\delta$-groups and
increase the efficiency of the broadcast payloads.  As pointed out in
\cite{delta-crdts}, ''\textit{a full state can be seen as a special (extreme)
  of a delta-group}''.

In the vanilla form of $\delta$-CRDTs, however, consistency is delayed ad-infinitum
when a message is lost and the per-object causal consistency property of
state-based CRDTs is lost.  These issues can
be addressed with an additional anti-entropy algorithm that groups, sorts,
tracks delivery and re-sends missing deltas, as presented in
\cite{delta-crdts}, but in the case of $\delta$-state-Merkle-CRDTs, 
the anti-entropy algorithm and any causal
information attached to the original objects would not be necessary.  In
essence, this approach brings $\delta$-state Merkle-CRDTs closer to their
operation-based counterpart.

\section{Limits and Optimizations of Merkle-CRDTs}\label{limits-optimisations}

\subsection{Limitations of Merkle-CRDTs}

We have so far focused on explaining the different qualities that
Merkle-CRDTs provide when compared to traditional CRDT approaches, but we must also
highlight what intrinsic and practical limitations they bring.

\paragraph{Ever-growing DAG-Size}
The most obvious consequence of Merkle-CRDTs is that, while CRDTs normally
merge, apply, consolidate and discard broadcast objects, Merkle-CRDTs build a
permanent Merkle-DAG which must be stored and is ever-growing. As we have seen, this
provides a number of advantageous properties, but also comes with some
implications:

\begin{itemize}
\item The size of the DAG might grow larger than acceptable. The rate of
  growth will depend on the number of the events and the size of the payloads.
  This is very similar to how blockhains grow to large sizes in
  time\footnote{Bitcoin chain uses more than 220GB and Ethereum (Parity) more
    than 165GB as of this writing.}. 
   For example, in a non-batched implementation,
 every insertion to a CRDT key-value store implementation will result in a new 
 DAG-node. Thus, for every key we will consume additional amount of space which,
 in the case of small objects, will likely be larger than the original object itself. 
 This is especially problematic when the
 actual state might be much smaller, for example when repeatedly updating a single key
 in a database. 
In some cases, it might be possible to express the state as a compact of the result of all the Merkle-CRDT operations,
  but this brings us to the next point.

\item If replicas store the Merkle-DAG only, knowing that the full state can be
  rebuilt from it (and thus saving that space), starting replicas with very
  large Merkle-DAGs might be especially slow since they will need to reprocess
  the full DAG, even when available locally. If not, there will be redundant
  information stored in both the resulting state and in the Merkle-DAG.

\item Merkle-CRDT syncs from scratch are possible and natural to the system
  when a new replica joins. However, Merkle-DAGs are not only ever-growing
  but also tend to be deep and thin\footnote{The Merkle-DAGs will be thin in
    the absence of many concurrent events, or have a high branching factor
    otherwise. In both cases, branches are consolidated every time a new event
    is issued from a replica, thus creating \textit{thin waists} in the
    DAG. }.  A new replica will learn the root CID from a broadcast operation
  and will need to resolve the full DAG from it. Because of the thinness, it
  will not be possible to fetch several branches in parallel. Cold-syncs may
  take significantly longer than it would take to ship a snapshot, thus
  rendering this embedded property of Merkle-DAGs of little value.
\end{itemize}

Very large DAGs and slow syncs are not a problem in some scenarios and can be
seen as an acceptable trade-off, but do highlight the need of exploring
``garbage collection" and DAG compaction mechanisms.

\paragraph{Merkle-Clock sorting}
Merging two Merkle-Clocks requires comparing them to see if they are included
in one another and finding differences. This may be a costly operation if DAGs
have diverged significantly (or long ago).

\paragraph{DAG-Syncer latency}
Replicas rely on a DAG-Syncer component to fetch and provide nodes from and to
the messaging layer. As mentioned earlier, Merkle-CRDTs are agnostic to the
mechanism used to synchronise messages (e.g., DHT or PubSub), but unless
chosen carefully, this mechanism might introduce sync delay. Depending on
application requirements, this delay might or might not be acceptable.

The practical impact of these limitations depends on the requirements of the
application. In particular, when thinking about adopting Merkle-CRDTs, users
should consider whether Merkle-CRDTs are the best approach in terms of: i)
Node count vs. state-size, ii) Time to cold-sync, iii) Update propagation
latency, iv) Expected total number of replicas, v) Expected replica-set
modifications (joins and departures), vi) Expected volume of concurrent
events.

\subsection{Optimizing Merkle-CRDTs}

\paragraph{Delayed DAG nodes}
In scenarios where replicas issue frequent updates, we can group multiple
payloads before issuing a single node containing all of them. It is clear that
this approach will bring some benefits, which however, comes with trade-offs:
updates are not immediately sent out and will, therefore, take longer to
propagate.

\paragraph{Quick Merkle-DAG inclusion check}M
erging the local replica DAGs with a remote one requires checking if one DAG
includes the other. It is possible but inefficient to do so by walking down
the first DAG looking for a node CID that matches the root of the second.
Storing the CIDs of the local DAG in a key-value store that can quickly check
whether a CID is part of the local DAG or not makes things significantly
easier\footnote{Fast key-value stores, such as in-memory ones, will normally
  pay a high memory footprint penalty, while disk-backed ones will be
  slower.}.  When walking the remote DAG to check for inclusion of the local
DAG, the CIDs of the children of any of its nodes can be checked to see if
they are part of the local DAG in which case their branches can be
conveniently pruned. This implies, however, that the implementation must be
aware and have access to the local storage system for nodes. The DAG-Syncer,
as currently defined, cannot differentiate between nodes available locally or
remotely.  Bloom filters, caches and some data structures can also improve
efficiency, but they are usually part of the chosen storage backend.

A similar effect can be achieved by embedding version vectors in the
payloads, as long as the application can tolerate the constraints they
impose. Comparing version vectors between payloads is an inclusion check
without the need to perform a DAG-walking.

\paragraph{Broadcast payload adjustments:}
Our standard approach reduces the size of the broadcasts by including only the
CID of the new roots. Publishing mechanisms are complex enough and always
benefit from smaller payloads.

However in some systems it may be beneficial to send new Merkle-DAG nodes
directly as part of the broadcast payloads. Replicas that are offline or
dropped messages will recover when they receive a future update and complete
their DAGs, so this has no effects in that regard. Broadcasting the payloads
(assuming they are small enough) will likely reduce the latency of the
propagation of changes in the system.

\paragraph{Reducing the Merkle-DAG node size}
We can attempt to reduce the size of the payloads as much as possible by
compressing and removing redundant information not required by the CRDT
itself. For example, instead of signing the CRDT payloads to ensure that they
come from a trusted replica, we can sign the broadcast messages, thus leaving
signatures out of the Merkle-DAG.

Another option is to make the payload (or parts of it) CIDs to reference the
actual contents. If the payloads are big, this will greatly reduce the size of
the Merkle-DAG and may increase the efficiency of the DAG fetching.  This is
especially relevant when some of the payloads are identical and can be
de-duplicated, or when it is possible to access part of the data
opportunistically.

\paragraph{Additional pointers in nodes:}
One of the ways to work around the thin-DAG problem is to regularly introduce
references to deeper parts of the DAG when issuing new nodes. This method is
basically adding extra children to nodes. It allows more parallelism when
fetching missing parts of the DAG by being able to jump to other
sections of it, resulting in much faster traversals. The actual number
of extra links and their destination will depend on the needs of the application.


The above recommendations should be considered in any Merkle-CRDT
implementation as they can provide significant advantages over the
un-optimized version described previously. Which optimizations fit to which
implementation is largely application-specific. We leave the topics of DAG
compaction and garbage collection for future work, although we intuitively
note that discarding parts of the Merkle-DAG should not be attempted before
making sure that every replica is aware of them.  This, in turn, requires
either having knowledge of the current replica-set or using an external source
of truth (i.e. a blockchain), a system constraint that we did not have before.

\section{Conclusion}

In this paper we approached Merkle-DAGs as causality-encoding structures with
self-verification and efficient syncing properties. This led us to introduce
the concept of \emph{Merkle-Clock}, demonstrating that they can be described
as a state-based CRDT which, announced with a \emph{Broadcaster} component and
fetched with a \emph{DAG-Syncer} facility, converges in all replicas.

We then presented \emph{Merkle-CRDTs} as Merkle-Clocks with CRDT payloads. We
showed how Merkle-CRDTs work with almost no messaging layer guarantees and
no constraints on the replica-set, which can be dynamic and unknown, while providing
per-object causal consistency. Merkle-CRDTs are widely used in the IPFS ecosystem for database logging operations\footnote{https://github.com/orbitdb/ipfs-log}, in \texttt{OrbitDB}\footnote{https://github.com/orbitdb/orbit-db}, a distributed, P2P database, its serverleess application, \texttt{Orbit}\footnote{https://github.com/orbitdb/orbit}, as well as distributed, collaborative editing\footnote{https://github.com/peer-base/peer-pad} and mobile photo-sharing applications\footnote{textile.photos}.

Merkle-CRDTs are a marriage between traditional blockchains, which need
consensus to converge, and CRDTs, which converge by design, and thus inherit
positive and negative aspects from both worlds. With this work, we hope to
have set a good foundation for further research on the topic.

\appendix

\section{Related work in the IPFS Ecosystem}

Merkle-CRDTs are very intuitive, even if they were not formalized before, and
rely on well-known and widely used properties of Merkle-DAGs. Several projects
in the IPFS ecosystem already use them\footnote{The dynamic data and
  capabilities working group has started many discussions on the topic:
  \url{https://github.com/ipfs/dynamic-data-and-capabilities}.}, all embedding
operation-based CRDTs in Merkle-DAGs:

\begin{itemize}

\item \texttt{ipfs-log}\footnote{\url{https://github.com/orbitdb/ipfs-log}}
  is, to our knowledge, the first existing instance of a Merkle-CRDT as
  described here. It implements an operation-based, append-only log CRDT
  (similar to a grow-only set).

\item
  \texttt{ipfs-hyperlog}\footnote{\url{https://github.com/noffle/ipfs-hyperlog}}
  is utility to build and replicate Merkle DAGs.

\item \texttt{Orbit DB}\footnote{\url{https://github.com/orbitdb/orbit-db}} is
  a distributed, peer-to-peer database. It uses \texttt{ipfs-log} and other
  CRDTs for different data models. It is used to build
  \texttt{Orbit}\footnote{\url{https://github.com/orbitdb/orbit}}, a
  distributed, serverless chat application.

\item \texttt{Tevere}\footnote{\url{https://github.com/ipfs-shipyard/tevere}}
  is an operation-based Merkle-CRDT key-value store.

\item
\texttt{peer-crdt}\footnote{\url{https://github.com/ipfs-shipyard/peer-crdt}}
and
\texttt{peer-crdt-ipfs}\footnote{\url{https://github.com/ipfs-shipyard/peer-crdt-ipfs}}
provide a generalistic operation Merkle-CRDT implementations of several CRDTs:
counters, sets, arrays, registers and text (as well as composable CRDTs).

\item
\texttt{versidag}\footnote{\url{https://github.com/ipfs/dynamic-data-and-capabilities/issues/50}}
is a proposed linked log with conflict resolution to store version
information, similar to \texttt{ipfs-log}.

\item
  \texttt{PeerPad}\footnote{\url{https://github.com/ipfs-shipyard/peer-pad}}
  is a real-time collaborative text editor based on \texttt{peer-crdt} and
  $\delta$-CRDTs.

\item
  \texttt{Textile.photos}\footnote{\url{https://www.textile.photos/}} is a
  mobile, decentralized digital wallet for photos. Textile Threads (v1) \cite{textile-threads}
  allow a group of users to share photos without a central database and
  are based on Merkle-CRDTs.

\item
  \texttt{go-ds-crdt}\footnote{\url{https://github.com/ipfs/go-ds-crdt}} is a
  key-value distributed datastore implementation in Go using $\delta$-state
  Merkle-CRDTs. It is used by IPFS Cluster\footnote{\url{https://cluster.ipfs.io}}.

\end{itemize}

%



\bibliographystyle{plain}
\bibliography{paper}
\balance

\end{document}